\documentclass[12pt]{cernprep}
\usepackage{graphicx}
\usepackage{fancyhdr}
\usepackage{subfigure}
\begin{document}

\begin{titlepage}
\docnum{CERN--PH--EP/2008--021}
\date{4 December 2008}

\title{\large {PDE-Foam - a probability-density estimation method\\
using self-adapting phase-space binning}} %% Paper title goes here


\begin{center}

%\vspace*{1cm} 

Dominik Dannheim$^{1ab}$,
Tancredi Carli$^1$, 
Alexander Voigt$^1$,\\
Karl-Johan Grahn$^2$, Peter Speckmayer$^{3c}$ 

\vspace*{5mm} 

$^1$~{\bf CERN, Geneva, Switzerland} \\
$^2$~{\bf KTH, Stockholm, Sweden} \\ 
$^3$~{\bf Technische Universit\"at, Wien, Austria} \\

\vspace*{5mm}

\begin{abstract}
Probability Density Estimation (PDE) is a multivariate discrimination
technique based on sampling
signal and background densities defined by event samples from data or Monte-Carlo (MC) simulations
in a multi-dimensional phase space. In this paper, we present a
modification of the PDE method that uses a self-adapting binning
method to divide the multi-dimensional phase space in a finite number
of hyper-rectangles (cells). The binning algorithm adjusts the size
and position of a predefined number of cells inside the
multi-dimensional phase space, minimising the variance of the signal
and background densities inside the cells. The implementation of the
binning algorithm (\texttt{PDE-Foam}) is based on the MC event-generation package \texttt{Foam}. 
We present performance results for representative examples (toy
models) and discuss the dependence of the obtained results on the
choice of parameters. The new \texttt{PDE-Foam} shows improved classification
capability for small training samples and reduced classification time
compared to the original PDE method based on range searching.
\end{abstract}

\submitted{appeared in: Nucl. Inst. and Meth. A 606, 717 (2009)}
\end{center}

\vspace*{5mm}
\rule{0.9\textwidth}{0.2mm}

\begin{footnotesize}
$^a$~Corresponding author; e-mail: dominik.dannheim@cern.ch. 

$^b$~Now with Max-Planck-Institut f\"ur Physik, M\"unchen, Germany.

$^c$~Now with CERN.
\end{footnotesize}

\end{titlepage}

\newpage

\section{Introduction}
Multi-variate discrimination techniques are used in high energy physics to distinguish 
signal from background events based on a set of measured
characteristic observables.
The information
contained in the individual observables is combined into a single
``discriminant'' variable, on which then a cut is applied to separate signal from background.
For an introduction to 
multi-variate discrimination techniques see e.g. Ref.~\cite{bishop}.

Besides other approaches, non-parametric probability density
estimaton (PDE) methods are used. 
Non-parametric PDE methods calculate a discriminant for each event to be classified based on the 
density of signal and background events in the vicinity of its coordinate
in the multi-dimensional phase space. In the following we consider
only methods that sample the event densities with probe volumes of fixed size. 
These methods have been used e.g. in searches for new physics at the 
Tevatron~\cite{top-tevatron} and at HERA~\cite{instanton-hera,isotau-hera}
and for particle identification at the LHC~\cite{atlas-cscbook}. 

A PDE method based on range searching (\texttt{PDE-RS})~\cite{pders-carli-koblitz}
has been used successfully for classification problems in higher-dimensional observable 
spaces and with arbitrary correlations between the observables. 
Large samples of Monte-Carlo (MC) simulated signal and background training events are stored in binary-search trees. 
An efficient range-searching algorithm is used to sample the signal and background 
densities in small multi-dimensional boxes around the phase-space points to be classified.
It handles the involved statistical uncertainties in a transparent way
and has the size of the sampling box as the only free parameter.

An apparent limitation of \texttt{PDE-RS}, on the other hand, is the fact that
large signal and background training samples are required to densely populate the 
multi-dimensional phase space. This is of particular importance in
applications with many dimensions. 
Furthermore, these samples have to be accessible in the main memory
of the computer used for the classification and the classification time scales with the
number of training events like $T_{class}\propto N_{train}\cdot \log N_{train}$. 
Though adaptive resizing and
kernel-convolution mechanisms for the sampling box have been
implemented for \texttt{PDE-RS} in the 
toolkit for multi-variate data analysis with ROOT (TMVA)~\cite{tmva},
the geometry 
of the sampling box is always identical in all dimension and it is 
therefore not optimally adapted to cases where the
density distributions vary for the different dimensions involved.

In this paper, we propose an improvement of the original
\texttt{PDE-RS} method~\cite{pders-carli-koblitz} that reduces the sensitivity to statistical fluctuations of the training samples and
results in a very fast and memory-efficient classification phase, independent of the size of the 
training samples. A self-adapting binning method is used to divide the multi-dimensional phase space in 
a finite number of hyper-rectangles (boxes). 
Only the binned density information is preserved in binary trees after the training phase, allowing for a very 
fast and memory-efficient classification of events. The implementation of the binning algorithm (\texttt{PDE-Foam})
is based on the MC event-generation package \texttt{Foam}~\cite{foam}.

\section{Probability Density Estimation}

PDE methods are based on the assumption that the probability for an event $\textbf{x}$ 
(characterised by a set of $d$ observables $x_1, ..., x_d$) to belong to the
signal class is given as a uniformly continuous function
$P(\textbf{x})$. According to Bayes' theorem, $P(\textbf{x})$ is
derived from the probability density functions for signal and
background, $\rho _s(\textbf{x})$ and $\rho _b(\textbf{x})$, and from
the a-priori probabilities $p_s$ and $p_b$ for an event to be of class signal or
background, respectively, as
\begin{equation}
P(\textbf{x})=\frac{\rho _s(\textbf{x})p_s}{\rho _s(\textbf{x})p_s+\rho _b(\textbf{x})p_b}.
\end{equation}
The signal probability $P(\textbf{x})$ is a monotonously rising
function of the \emph{Discriminant}
$D(\textbf{x})$:
\begin{equation}
D(\textbf{x})=\frac{\rho _s(\textbf{x})}{\rho _s(\textbf{x})+\rho
  _b(\textbf{x})}
\end{equation}
in the relevant range $0 \leq D(\textbf{x}) \leq 1$,
such that a cut on $D$(x) is always equivalent to a cut on $P$(x).
$P$(x) and hence $D$(x) are
the optimal discriminants according to the Neyman-Pearson 
lemma~\cite{neyman-pearson}.

Estimates $\rho' _s(\textbf{x})$ and $\rho' _b(\textbf{x})$ of the signal and
background probability density functions are obtained by
sampling the $d$-dimensional phase space with 
events of known type. Such events can be either obtained from MC simulations or by defining
data control samples in an appropriate way.
The estimated discriminant: 
\begin{equation}
D'(\textbf{x})=\frac{\rho' _s(\textbf{x})}{\rho' _s(\textbf{x})+\rho'
  _b(\textbf{x})})\approx D(\textbf{x})
\end{equation}
approximates the true discriminant $D(\textbf{x})$ for sufficiently
densely populated sampling space.

For any given combination of observables, $\textbf{x}$, the discriminant $D'(\textbf{x})$ assigns a single
value, which allows to discriminate background from signal events. In the framework of a physics
analysis, a cut on a particular value of $D'$ is applied, depending on the required purity
and efficiency of the event selection.
The signal and background probability density functions
have to be approximated with sufficient accuracy. This poses
a challenging problem in particular for high-dimensional cases. 

A solution based on range searching (\texttt{PDE-RS}~\cite{pders-carli-koblitz})
counts the number of MC generated signal and background events in the vicinity of
each event to be classified. The discriminant $D'(\emph{x})$ is defined from the number
of signal events $n_s$ and the number of background events $n_b$ in a small volume $V(\textbf{x})$
around the point $\textbf{x}$:
\begin{equation}
D'(\textbf{x})=\frac{n_s}{n_s+c\cdot n_b}.
\end{equation}
The normalisation constant $c$ has to be chosen such that the total
number of simulated signal events, $N_s$,
is equal to $c$ times the total number of background events, $N_b$:
\begin{equation}
\label{norm_const}
N_s=c\cdot N_b
\end{equation}

The statistical uncertainty on the value of the discriminant $D$ is obtained from a propagation
of the uncertainty on the number of events contained in the counting volume:
\begin{equation}
\sigma_D'(x)=\sqrt{ \left( \frac{c\cdot n_b}{(n_s+c\cdot n_b ) ^2}\sigma_{n_s}\right)^2 
  + \left( \frac{c\cdot n_s}{(n_s+c\cdot n_b ) ^2}\sigma_{n_b}\right)^2},
\label{eq:stat_error}
\end{equation}
where $\sigma_{n_s}$ and $\sigma_{n_b}$ are the statistical uncertainties of the number of
signal and background events respectively.

\begin{figure}[htp!]
\centering
   \subfigure[Discriminant distribution]{%
      \label{fig:discr}
      \includegraphics[width=0.48\textwidth]{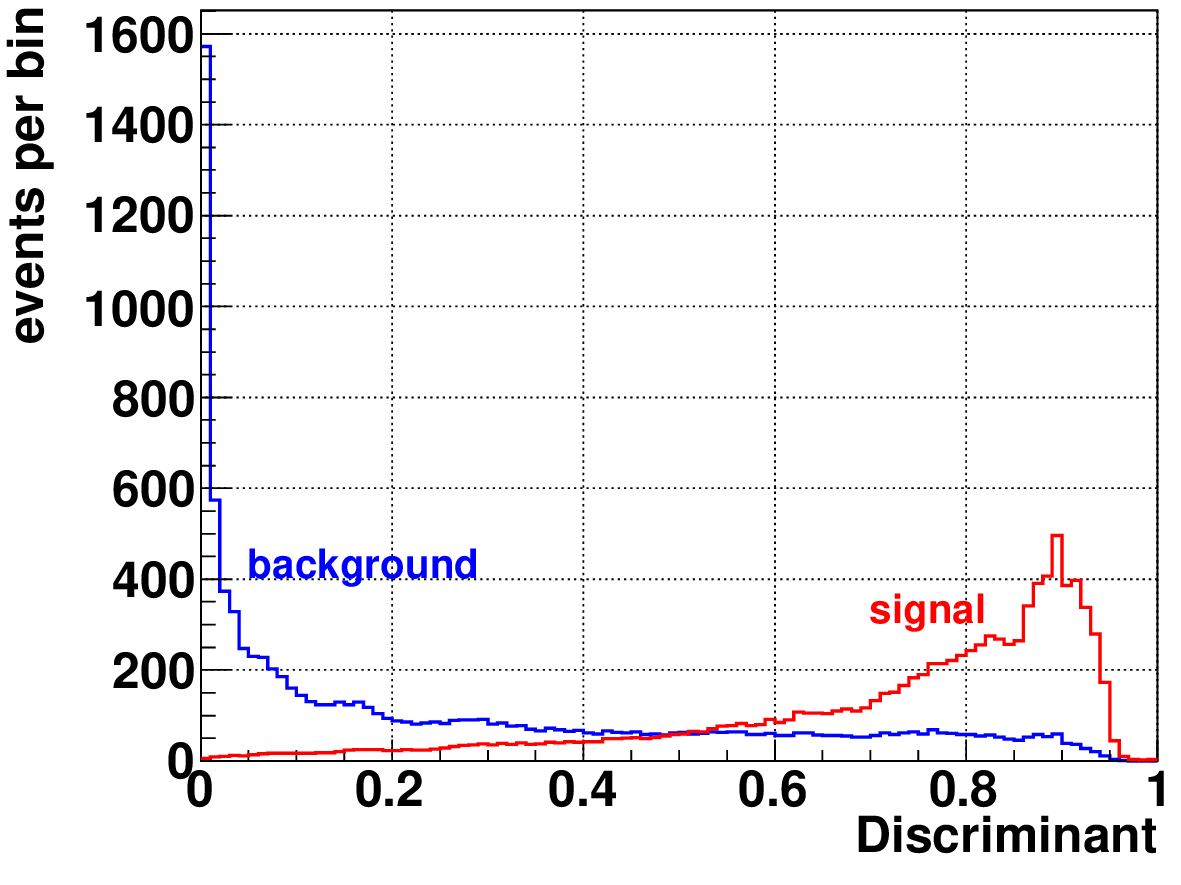}}
   \subfigure[Receiver Operating Characteristic]{%
      \label{fig:roc}
      \includegraphics[width=0.48\textwidth]{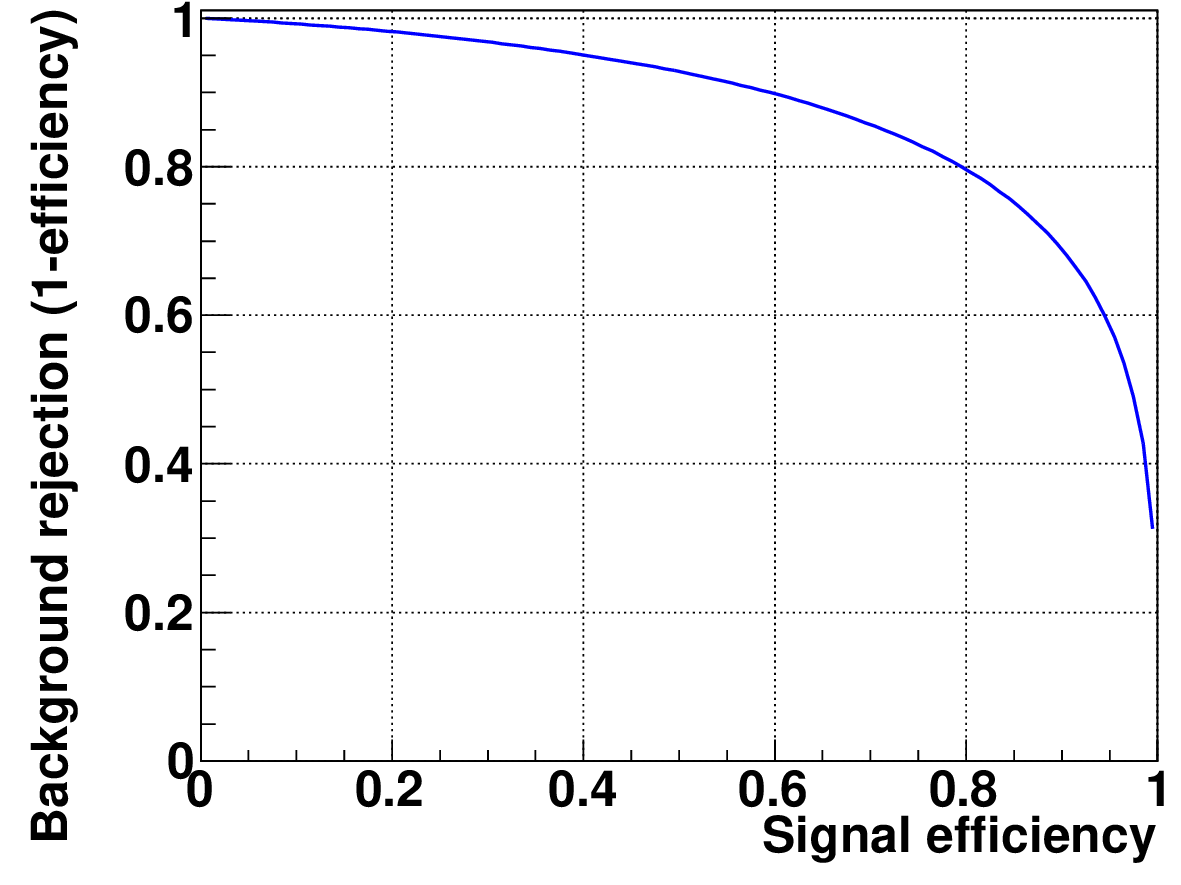}}
\caption[.]{(a) Distribution of the discriminant for an arbitrary example of signal and
  background MC events. (b) ROC curve for the same example. The area under the curve characterises the
  performance of the estimator.}
\label{fig:discr_roc}
\end{figure}

$D'(\textbf{x})$ provides a good estimate of
$D(\textbf{x})$ for sufficiently small probe volumes $V(\textbf{x})$ and large numbers of MC simulated
sample events. Figure \ref{fig:discr} shows the distribution of the discriminant $D'(x)$ for signal and background
testing events of an arbitrary example. The discriminant takes values between 0 and 1. 
Most signal events are found at large values of $D'$, while the distribution 
for background events peaks at small values of $D'$. A given cut value $D_c$ results in an efficiency for signal
and background testing samples, $\epsilon_{s}(D_c)$ and $\epsilon_{b}(D_c)$. Figure~\ref{fig:roc} shows
the relation between signal efficiency $\epsilon_{s}$ and background rejection $1-\epsilon_{b}$ 
when scanning all values of $D_c$ between 0 and 1. The area under this so called \emph{Receiver operating
characteristic} (ROC) is a measure of the average estimator performance. A value of 0.5 is obtained for a random classification
of signal and background. A value of 1.0 is obtained for a perfect discrimination between signal and background.
For the example shown in Fig.~\ref{fig:discr_roc}, the obtained ROC
value is 0.88. 

The optimal performance
measure for a particular practical application has to be chosen according to the desired 
balance between purity and efficiency. In high energy physics, MVA
methods are often used in searches for rare events, where a small
signal is overwhelmed by background. In such cases, the signal
efficiency for a given large value of the background rejection is a
more relevant performance measure than the overall area of the ROC
curve. In addition to the ROC area, one therefore often quotes 
$\epsilon_{S,R=99\%}$, the signal efficiency at a background rejection of 99\%.
For the example shown in Fig.~\ref{fig:discr_roc}, the resulting value is $\epsilon_{S,R=99\%}=13.0\%$.

\section{Adaptive phase-space binning with \texttt{Foam}}
In the following we propose an alternative method to calculate the discriminant $D(\textbf{x})$
based on a binned sampling of the phase space. 

Simple binning methods 
often suffer from excessive memory consumption and lack of statistical accuracy, since the number 
of bins increases as $m^d$, where $m$ is the number of bins per dimension. The bin size
has to be small enough to follow fine changes in the event distributions in phase-space
regions where signal and background overlap. This often leads to a large number of scarcely populated
bins. 
In many practical applications the phase space is effectively only populated in
a sub-space of lower dimensionality, since the \emph{intrinsic dimensionality} of the actual
problem is often reduced due to correlations among the observables. 

To overcome this problem, a self-adaptive binning method, called ``\texttt{PDE-Foam}'',
is used to project the information contained in the signal and background samples
into a grid of $d$-dimensional cells with non-equidistant cell boundaries, called 
the ``foam of cells''.

The method is based on an algorithm originally developed for the 
multi-dimensional general purpose MC event generator \texttt{Foam}~\cite{foam}.
For a given $d$-dimensional analytically known distribution, 
the \texttt{Foam} algorithm creates a hyper-rectangular "foam of cells", 
which is more dense around the peaks of the distribution and less dense in areas where
the distribution is only slowly varying.
The foam is iteratively produced 
using a binary-split algorithm for the cells acting on samplings of the input distribution
within the cell boundaries.
The number of cells is a predefined free parameter and a priori only limited by the amount of available 
computer memory. The optimal number of cells depends on the
statistical accuracy of the training samples.

\subsection{The \texttt{PDE-Foam} build-up algorithm}\label{subsec:foam-algorithm}
\label{foam-build-up-algorithm}
In the context of PDE, \texttt{Foam} has been adapted such that the splitting of cells is based on
an input distribution that is sampled from MC training events using
the \texttt{PDE-RS} method. The steering parameters introduced in the following are
summarised in Table~\ref{tab:parameters} and their usage and optimisation is discussed in 
section~\ref{sec:pdefoam-parameters}.

\begin{table}
\begin{center}
\begin{tabular}{|l|l|l|}
\hline
Parameter           & default & Description\\
\hline
\texttt{TailCut}    & 0.001 & Fraction of outlier events excluded
from the foam\\
\texttt{VolFrac}    & 1/30 & Fraction of foam volume used for sampling\\ 
                    &      & during the foam build-up phase\\
                    &       &  (1 is equivalent to the volume of entire foam)\\
\texttt{nActiveCells}        & 500  & Maximal number of active cells
that can be\\
                    &      & created during the foam build-up\\
\texttt{Nmin}                & 100   & Minimum number of events per cell \\
\texttt{nSampl}              & 2000  & Number of samplings per cell and cell-division\\
                    &       & step during the foam build-up phase\\
\texttt{nBin}                & 5     & Number of bins for edge histograms\\
\texttt{Kernel}              & None  & Used kernel estimator (None or Gauss)\\
\hline
\end{tabular}
\caption{Main \texttt{PDE-Foam} parameters and their default values. The
  parameters and their optimisation are discussed in 
  sections \ref{subsec:foam-algorithm} and \ref{sec:pdefoam-parameters}.}
\label{tab:parameters}
\end{center}
\end{table}

The build-up of the foam starts with the creation of the base cell, 
which corresponds to a $d$-dimensional hyper-rectangle containing
all MC training events.

The coordinate system of the foam is normalised such that the base
cell extends from 0 to 1 in each dimension. The coordinates of the
events in the corresponding training samples are linearly transformed
into the coordinate system of the foam. Tails of the input
distributions are removed from the base cell by an adjustable parameter
\texttt{TailCut}. An upper and a lower bound are determined for each
dimension such that on both sides of the corresponding one-dimensional 
distribution a fraction of \texttt{TailCut} of all events are 
excluded\footnote{
Note that, for the classification of events, it is guaranteed that
the foam has an infinite coverage: events outside the foam volume are
assigned to the cells with the smallest cartesian distance to the
event.}.

Starting from this base cell, a binary splitting algorithm iteratively
splits cells of the foam along hyperplanes until a predefined maximum
number of cells, \texttt{nActiveCells},
is reached.
The implementation is identical to the one of the original \texttt{Foam} 
code~\cite{foam}.
It minimises the relative variance of the
density $\sigma_{\rho}/\langle\rho\rangle$ across each
cell\footnote{The density $\rho$ is either defined as the sampled density of 
events of a given type or as the sampled density of the discriminant,
as will be discussed in the following section.}.  

For each cell a predefined number \texttt{nSampl} of
random points
uniformly distributed over the cell volume are generated. 
For each
of these points a small box of fixed size \texttt{VolFrac} centred around this point 
is considered to estimate the local event density of the corresponding training
sample as the number of training events contained in this box
divided by its volume. Events from neighbouring cells are also counted 
in cases where the sampling box extends beyond the cell boundaries. 
The obtained densities for all sampled points in the cell are
projected on the $d$ axes of the cell and the projected values are
filled in histograms with a predefined number of
bins, \texttt{nBin}.  

The cell to be
split next and the corresponding division edge (bin) for the split
are selected as the ones that have the largest relative variance.
After the split, the two new daughter cells become `active' cells and the
old mother cell remains in the binary tree, marked as being `inactive'.  
A detailed description
of the splitting algorithm and the \texttt{Foam} data structure can be found elsewhere~\cite{foam}. 

\begin{figure}[htp!]
\centering
   \subfigure[Input density]{%
      \label{fig:2d_gauss}
      \includegraphics[width=0.48\textwidth]{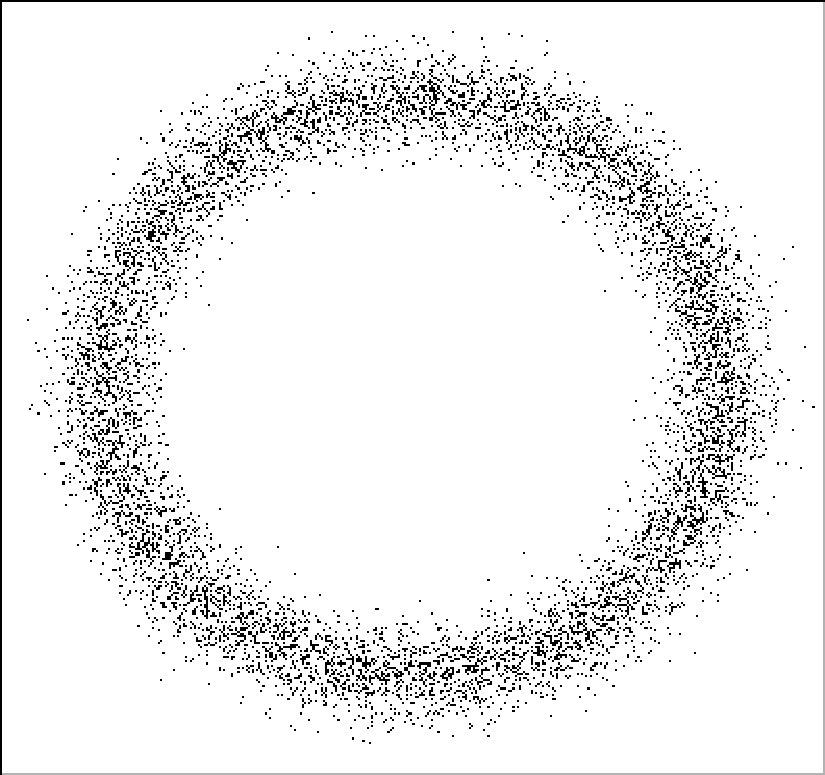}}
   \subfigure[Foam representation]{%
      \label{fig:foam_density_2000c}
      \includegraphics[width=0.48\textwidth]{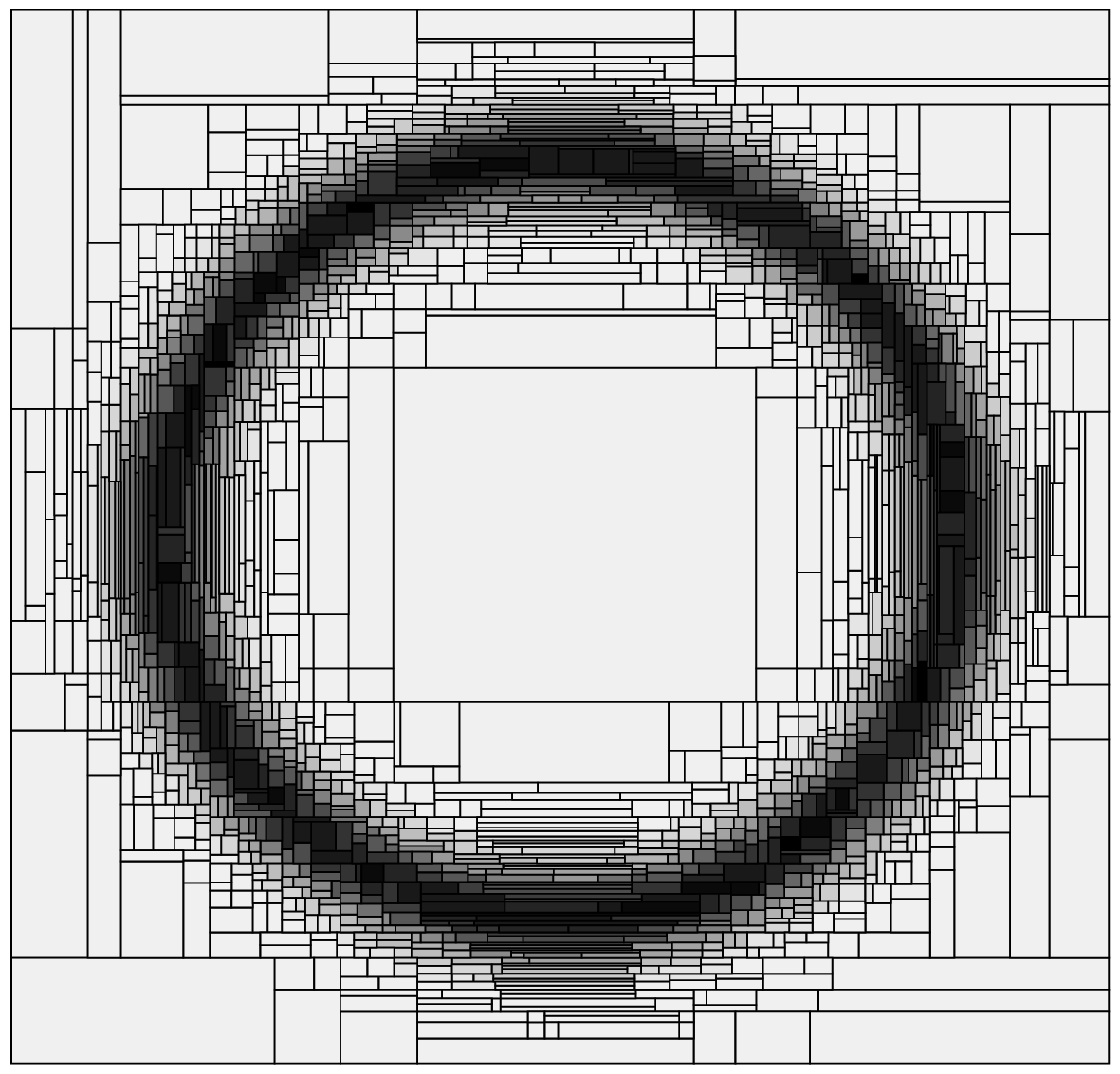}}
\caption[.]{(a) Two-dimensional Gaussian-ring distribution sampled from
500000 events. (b) Foam representation with 
2000 active cells. The level of grey
    indicates the event density inside the corresponding cell.}
\label{fig:foam_example}
\end{figure}

The
geometry of the final foam reflects the distribution of the
training sample: phase-space regions where the density is approximately
constant are combined in large cells, while in regions with large
gradients in density many small cells are created.
Figure~\ref{fig:2d_gauss} shows a two-dimensional Gaussian-ring 
distribution\footnote{The definition of this 
Gaussian-ring distribution corresponds to the signal distribution of the example 
``Highly Correlated Observables'' defined and discussed in Ref.~\cite{pders-carli-koblitz}.
The events are distributed uniformly in the azimuth angle and according to
a Gauss distribution in the radial coordinate, with a mean radius of 0.3
for both signal and background and a width
of 0.025 (0.05) for signal (background).} and
Fig.~\ref{fig:foam_density_2000c} 
shows a graphical representation 
of the resulting foam with 2000 active cells.
Each cell contains the number of events from the input distribution belonging to the
volume of the cell.
The foam consists of only a few large and sparsely populated cells in the centre and corner regions
of the two-dimensional plane, where the gradient of the Gaussian radial 
component of the distribution is small.
Close to the centre of the ring, however, where the radial component 
of the distribution has a steep gradient, the foam consists of many 
small and densely populated cells. This example is particularly challenging for
the foam algorithm, as the rectangular geometry of the foam cells does not match
the angular symmetry of the example\footnote{The original \texttt{Foam} implementation~\cite{foam}
has an option to define cells with simplicial shape, which however we
do not consider here.}.

The foam structure is formally equivalent to a decision
tree~\cite{decision-tree}. The cut values of the decision tree
correspond to the cell-splitting boundaries stored in the binary tree
representing the foam. Optimisation of the decision tree
(e.g. boosting) is replaced in case of \texttt{PDE-Foam} by the sampling and
minimisation algorithm described above. 

\subsection{Foam Application for MVA}
In order to use the \texttt{Foam} for MVA, two different concepts have been 
implemented:

\begin{enumerate}
\item \emph{Separate signal and background foams}

During the \emph{training phase} two separate foams are created:
one for signal and one for background events. The splitting of cells
is based on the corresponding event densities. The number of signal 
(background) events contained in each cell of the final signal (background)
foam is stored with the corresponding cell. During the \emph{classification phase}
the value of the discriminant for a given event $\textbf{x}$ is calculated based
on the number of events contained in the corresponding cells:
\begin{equation}
D(\textbf{x})=\frac{n_i/V_i}{n_i/V_i+c\cdot n_j/V_j},
\end{equation}
where $n_i$ and $n_j$ are the number of events contained in cell $i$ 
of the signal foam and cell $j$ of the background foam, respectively, and
$V_i$ and $V_j$ are the cartesian volumes of cell $i$ and $j$, respectively.

The statistical uncertainty on the discriminant $\sigma_D(\textbf{x})$ is obtained in analogy to
eq.~\ref{eq:stat_error}.

\item \emph{One foam for discriminant distribution}

During the \emph{training phase} one foam is created containing the
distribution of the discriminant. The splitting of cells is based
on the sampled discriminant distribution calculated according to
the \texttt{PDE-RS} approach. Each cell $i$ contains a discriminant value $D_i$ calculated
as:
\begin{equation}
D_i=\frac{n_{s,i}}{n_{s,i}+c\cdot n_{b,i}},
\end{equation}
where $n_{s,i}$ ($n_{b,i}$) are the number of signal (background) events
contained in cell $i$. The statistical uncertainty on the discriminant obtained in
analogy to eq. \ref{eq:stat_error} is also stored with each cell. During the 
\emph{classification phase} the value of the discriminant for a given event
$x$ from and independent testing sample and its statistical uncertainty are 
retrieved from the corresponding cell $i$.
\end{enumerate}

For the same number of total foam cells, the performance of the two
implementations was found to be similar.

\subsection{Foam classification example}
\begin{figure}[htp!]
\centering
   \subfigure[Discriminant (100 cells)]{%
      \label{fig:discr_foam_100c}
      \includegraphics[width=0.48\textwidth]{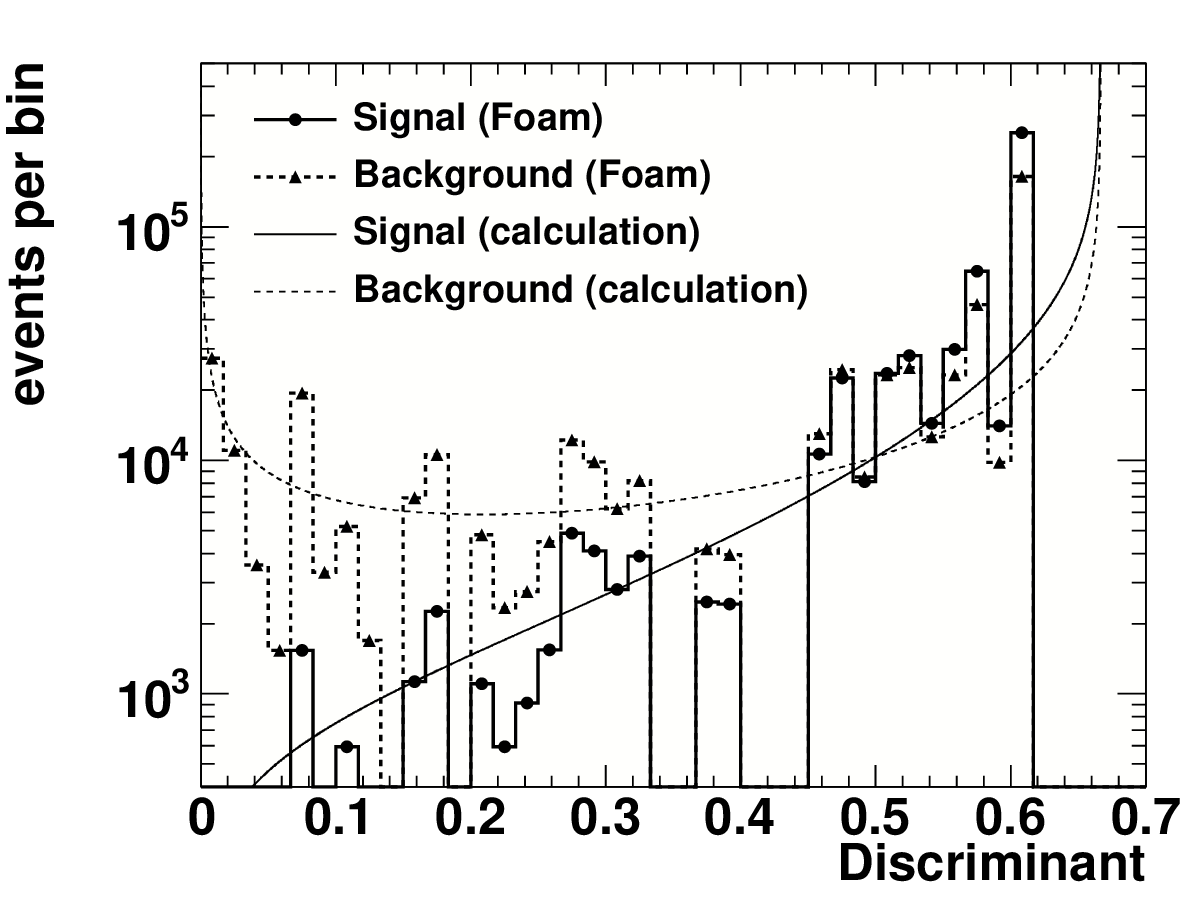}}
   \subfigure[Discriminant (500 cells)]{%
      \label{fig:discr_foam_500c}
      \includegraphics[width=0.48\textwidth]{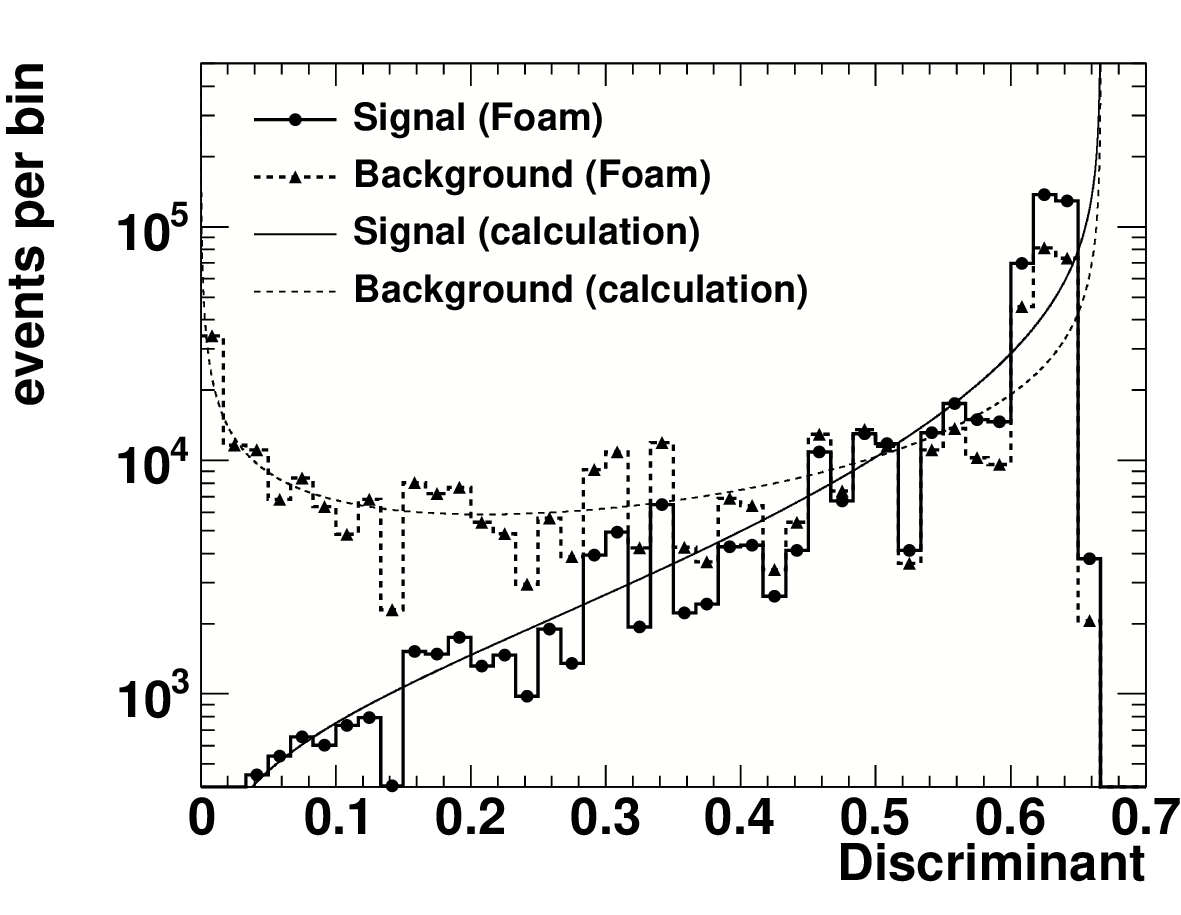}}
   \subfigure[Discriminant (2000 cells)]{%
      \label{fig:discr_foam_2000c}
      \includegraphics[width=0.48\textwidth]{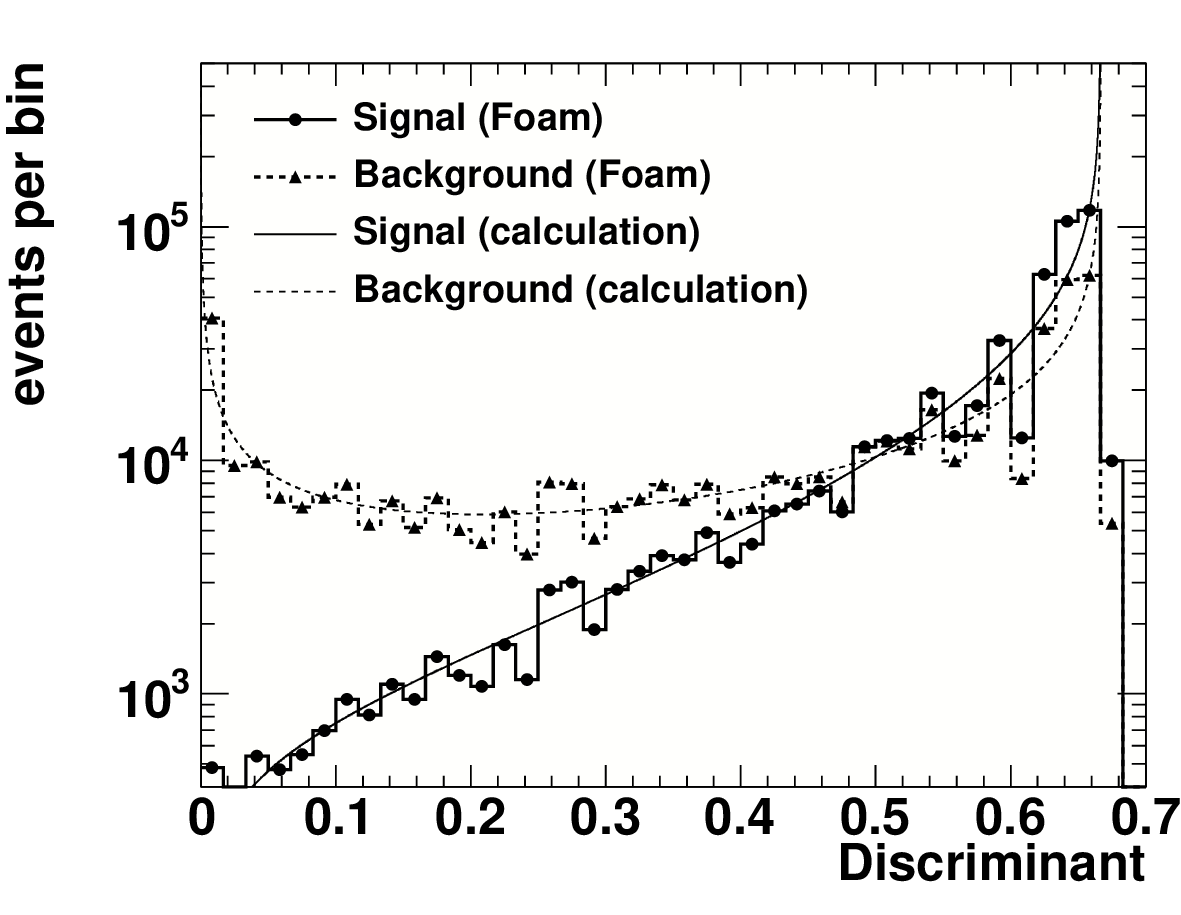}}
   \subfigure[Receiver Operating Characteristics]{%
      \label{fig:roc_foam}
      \includegraphics[width=0.48\textwidth]{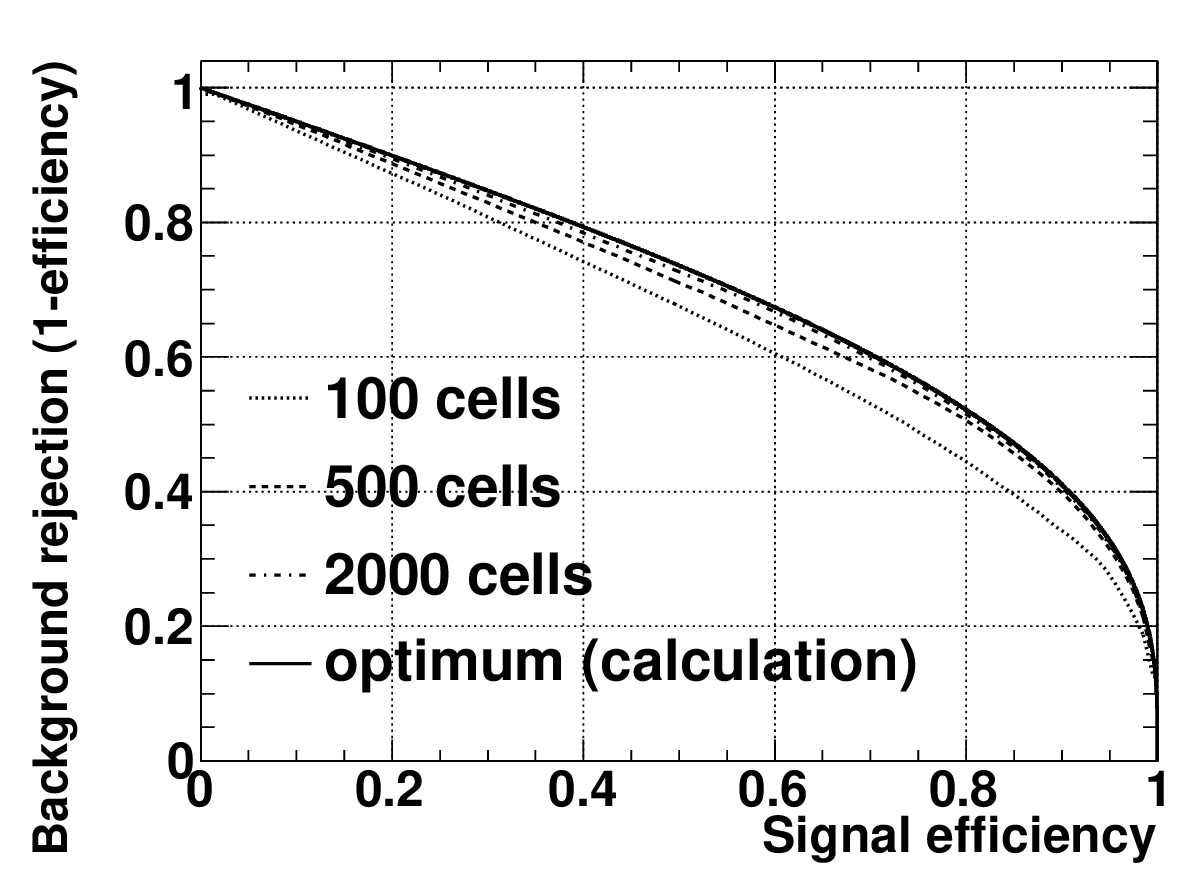}}
\caption[.]{Distribution of the discriminant for signal and background
  events of the Gaussian-ring example using 100 (a), 500 (b)
  and 2000 (c) active cells, respectively. The curves overlaying the
  histograms show the distributions corresponding to an analytical calculation. The 
  resulting ROC curves for the same example are shown in (d), together
  with the optimal curve obtained from the analytical calculation.}
\label{fig:discr_roc_foam}
\end{figure}

Figures~\ref{fig:discr_roc_foam}(a)-(c) show the distribution of the discriminant
for 500000 signal and 500000 background testing events of the Gaussian-ring
example introduced above. The classification is performed with single
discriminant foams of 100, 500 and 2000 active cells,
respectively. The foams are created using 200000 signal and 200000 background training events. 
Besides the histograms for the events classified with \texttt{PDE-Foam}, also the
corresponding curves for an analytical calculation are shown in the Figures. 

For this example, both signal and background input distribution
peak at the same values and are only distinguished by their different
width. Therefore the resulting discriminant distributions for signal
and background show a large overlap and poor separation. 

The peaked structure of the histograms reflects the finite granularity of the
foams. The distributions become smoother and approach the analytically
calculated curves with increasing number of foam cells.

Figure~\ref{fig:roc_foam} shows the resulting ROC curves for the
same example. The curve for the theoretical optimum obtained from the analytical
calculation is also shown. The area under the curve increases with increasing
number of foam cells from 0.655 (100 cells) to 0.699 (2000 cells). 
The theoretical optimum corresponds to an area of 0.705.
For a background rejection of 99\%, the respective signal-efficiency
values are between 1.5\% (100 cells) and 1.85\% (2000 cells) for the
classification with \texttt{PDE-Foam} and 2.0\% for the theoretical optimum.

\section{Implementation}

The foam algorithm is implemented within the TMVA framework~\cite{tmva}
as method \texttt{PDE-Foam}. The core foam functionality is inherited from the TFoam
class included in ROOT~\cite{root}. Parameter steering, classification output
and persistency mechanism follow TMVA standards, thus allowing to use
it like other MVA methods implemented in TMVA and to compare the performance. 
The initial binary trees, which contain the training events, needed to
evaluate the densities for the foam build-up based on the \texttt{PDE-RS} method, 
are discarded after the training phase.
  
The memory consumption for the foam is $160$ bytes
per foam cell plus an overhead of $1.4$ kbytes for the \texttt{PDE-Foam} object
on a $64$-bit architecture.  Note that in the foam all cells created
during the growing phase are stored within a binary tree
structure. Cells which have been split are marked as inactive and
remain empty. To reduce memory consumption, the geometry of a cell is
not stored with the cell, but rather obtained recursively from the
information about the division edge of the corresponding mother
cell. This way only two short integer numbers per cell contain the
information about the entire foam geometry: the division coordinate
and the bin number of the division edge. 

The foam object can be stored in XML or ROOT
format. A projection method is available for visible inspection. 

\subsection{\texttt{PDE-Foam} parameters} \label{sec:pdefoam-parameters}
Table~\ref{tab:parameters} summarises the main \texttt{PDE-Foam} parameters
that can be set by the user together with their default values.
Optimisation of these parameters is needed to reach optimal classification performance.
In the following, we discuss the dependence of the foam performance on the choice of parameters 
for some representative examples.

\subsubsection{Size of sampling box}
The size of the box used for the phase-space sampling is a common parameter of both the
\texttt{PDE-Foam} method and the original \texttt{PDE-RS} method. In case of \texttt{PDE-Foam}, the box size is only
relevant for the density sampling during the training phase, while for \texttt{PDE-RS} the box size
is only used for the calculation of the discriminant during the classification phase.
A larger box leads to a reduced statistical uncertainty for small
training samples and to smoother sampling. 
A smaller box on the
other hand increases the sensitivity to statistical fluctuations in the training samples, but
for sufficiently large training samples it will result in a more precise local estimate 
of the sampled density. 

Besides affecting the estimator performance, the box size influences the 
training time in case of \texttt{PDE-Foam} and the classification time in case of \texttt{PDE-RS}.
A larger box increases the CPU time during sampling,
due to the larger number of nodes to be considered in the binary 
search~\cite{pders-carli-koblitz}.

In general, higher dimensional problems
require larger box sizes, due to the reduced average number of events
per box volume. For uniformly distributed variables, the volume size
containing a given number of events grows with the power of the number
of dimensions. To collect 10\% of the training events inside the
sampling volume for a case with 10 variables, a box with edge length 
of 80\% of the full range in each dimension is needed.

\begin{figure}[htp!]
\centering
   \subfigure[\texttt{PDE-Foam}]{%
      \label{fig:volfrac}
      \includegraphics[width=0.48\textwidth]{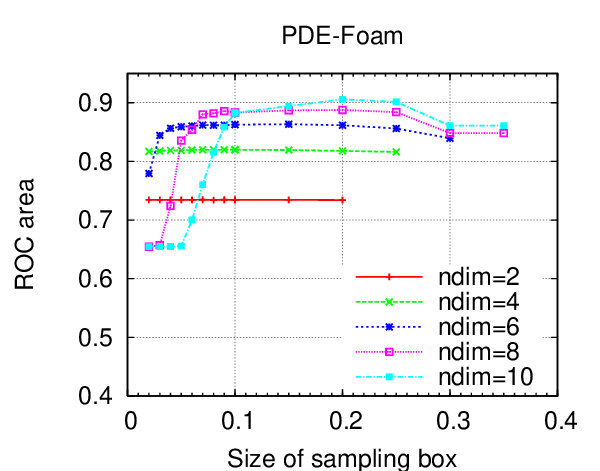}}
   \subfigure[\texttt{PDE-RS}]{%
      \label{fig:deltafrac}
      \includegraphics[width=0.48\textwidth]{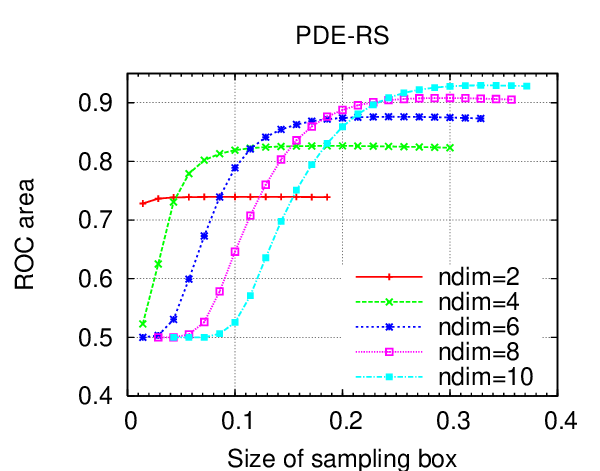}}
\caption[.]{Dependence of the estimator performance on the size of the sampling box and the
number of observables. The examples are constructed as Gaussian distributed
observables with shifted means for signal and background and with variable
number of observables.
50000 signal and 50000 background events each were used for the training phase. 
Shown are the performance for \texttt{PDE-Foam} (a) and for the original \texttt{PDE-RS} method (b).}
\label{fig:boxsize}
\end{figure}

Figure~\ref{fig:boxsize} shows the estimator performance, measured as the area under the
ROC curve, as function of the size of the sampling box and for examples with 2-10 observables
(=dimensions).
The examples are constructed as uncorrelated $n$-dimensional Gauss distributions with shifted means
and different widths for
signal and background\footnote{The values of all $n$ observables ($n=2..10$) are generated
from Gauss distributions with mean values $\bar{x}_{s}=5$ and $\bar{x}_{b}=4.5$
and widths $\sigma_{s}=0.75$ and $\sigma_{b}=1.0$ for signal and background,
respectively.}. 

The \texttt{PDE-Foam} performance (see Fig.~\ref{fig:volfrac}) is
compared to the performance of the original \texttt{PDE-RS} method (see Fig.~\ref{fig:deltafrac}). 
In both cases 50000 signal and 50000 background
training events were used. 
For \texttt{PDE-Foam}, a target value of 1000 active cells was selected 
and a cut on the minimum number of events per cell 
of 100 (cf. discussion in section~\ref{sec:nmin}) was applied during the foam build-up.
Here and in the following, the performance values have been obtained from independent 
testing samples of 500000 signal and 500000 background events.

The performance increases for both methods with the number of 
observables and with the size of the sampling box. It reaches a
maximum for both methods, after which 
it drops again slightly with further increasing box size, due to the less
precise local estimate of the larger boxes.
\texttt{PDE-Foam} is less sensitive to statistical fluctuations in the training
samples, due to the additional averaging stage during the density
sampling inside the cells. Therefore the \texttt{PDE-Foam} performance reaches
the optimum for smaller sampling boxes and has 
a wider range of stable performance, compared to the original \texttt{PDE-RS} implementation. 
For a small number of up to approximately four observables, there is almost no visible dependency of
the performance on the box size. The default box size of 1/30 gives close to
optimal results up to approximately five observables for this example.
For the original \texttt{PDE-RS} method,
on the other hand, a more careful optimisation of the box size is required, as the box size
for optimal performance depends strongly on the number of dimensions and the convergence
towards optimal performance is slower.

\subsubsection{Number of cells}
The target number of cells for the final foam is the main parameter impacting
the accuracy of the phase-space binning. An increased number of cells leads in general to
improved performance provided that sufficiently large training samples are available.
However, for an increasing number of cells with small training samples, 
the foam becomes more vulnerable to statistical fluctuations in the training samples in particular
in less populated regions of the phase space and 
the performance might drop when further increasing the target number of cells (overtraining). 
Both the training time and the memory needed to store the foam object
increase linearly with the number of cells, while the classification
time scales approximately as $T_{class}\propto N_{cells}\cdot \log N_{cells}$.

\begin{figure}[htp!]
\centering
\includegraphics[width=0.48\textwidth]{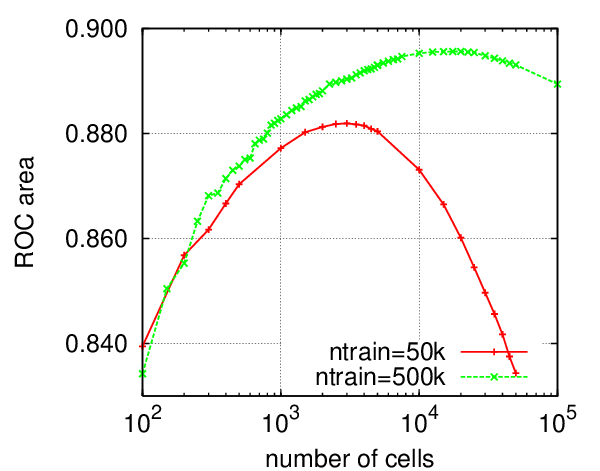}
\caption[.]{Dependence of the estimator performance on the number of active foam cells
for an example with five moderately correlated observables constructed
from Gaussian distributions and for two different training-sample sizes.}
\label{fig:ncells}
\end{figure}

Figure~\ref{fig:ncells} shows the dependence of the
estimator performance as function of the number of active cells for an example with five moderately
correlated observables constructed from Gaussian distributions for signal and 
background\footnote{The definition of the distributions
corresponds to the example ``High Dimensional Example''
defined and discussed in Ref.~\cite{pders-carli-koblitz}.}. The two curves
correspond to foams build-up
from small and large training samples, respectively. The small training sample consist of
50000 signal and 50000 background events, whereas the large training sample contain
500000 signal and 500000 background events. No restriction on the
number of events contained in each cell was applied.

As expected, the performance of the foams built from the large training sample 
exceeds the one of the foams based on the small training sample.
In case of the large training sample, the performance for this
particular example increases over
a wide range of number of cells and reaches its maximum for about 20000 cells, 
after which it drops due to the decrease in statistical precision resulting in overtraining.
For the small training sample, the maximum is already reached for foams with approximately
5000 cells and the drop in performance afterwards is steeper.

\subsubsection{Minimum number of events}
\label{sec:nmin}
The cell splitting algorithm assumes sufficient statistical accuracy of the sampled
density distributions in all cells. This might not be guaranteed in case of small
training samples, where cell splitting in scarcely populated phase-space regions can
lead to overtraining effects. Therefore cells should not be taken into account 
for further splitting, if the number of training events contained inside a cell is too small.

An adjustable parameter \texttt{Nmin} has been implemented, which sets the minimum number 
of events contained in any cell which is considered for further splitting. If the number
of events is below \texttt{Nmin}, the cell is not considered for further splitting. If no more
cells are available with sufficient number of events, the cell splitting stops, even
if the target number of cells is not yet reached. Note that the
\texttt{Nmin} requirement only affects the further splitting of
cells. Therefore it is possible to have cells containing less than
\texttt{Nmin} events in the final foam(s).

\begin{figure}[htp!]
  \centering
  \subfigure[Estimator performance]{%
    \label{fig:nminperf}
    \includegraphics[width=0.48\textwidth]{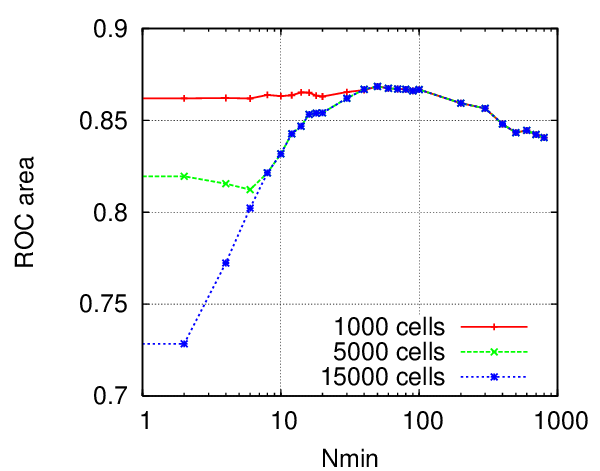}}
  \subfigure[Number of active cells in the final foam]{%
    \label{fig:nmincells}
    \includegraphics[width=0.48\textwidth]{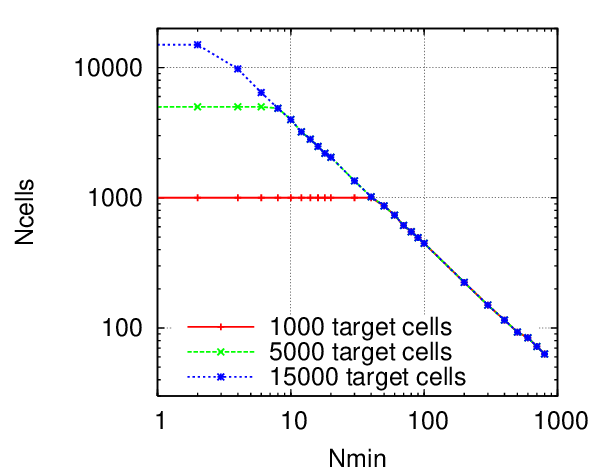}}
  \caption[.]{Estimator performance (a) and number of active cells in the final foam (b)
    as function of the cut value for the minimum number
    of events per cell. The data shown corresponds to the
    example with five moderately correlated observables constructed
    from Gaussian distributions. It
    was calculated using 10000 signal and background training events 
    each and target values of 1000, 5000 and 15000 active cells, respectively.}
  \label{fig:nmin}
\end{figure}

The cut on \texttt{Nmin} reduces 
the sensitivity to statistical fluctuations in the training samples and improves 
drastically the performance for small number of training events, as shown in Fig.~\ref{fig:nminperf}
for the example with 5 observables and only 10000 training signal and background events each.
Without the cut on \texttt{Nmin}, the foams with larger number of target cells suffer
from overtraining and show a significantly decreased performance.
Starting from a value of $\texttt{Nmin}\approx 8$, the effective number of final cells, as shown
in Fig.~\ref{fig:nmincells} 
is limited to a value below 10000 and therefore the performance curves for 10000 and 
30000 target cells become identical (Fig.~\ref{fig:nminperf}). 
For a value of $\texttt{Nmin}\approx 40$, this number drops
to 2000 cells, visible in both figures as the points where all three curves merge. For very large
values of \texttt{Nmin} the performance drops again, as the size of the few remaining cells
becomes too large.

The default value of $\texttt{Nmin}=100$ leads to a good performance
for most cases studied. It can be combined with a large target number of cells, as it limits
the effective number of cells sufficiently and thus avoids overtraining even for small training-sample 
sizes. For the example shown in Fig.~\ref{fig:nmin}, the value of $\texttt{Nmin}=100$ corresponds to
approximately 450 active cells in the final foams.

\subsubsection{Number of samplings}
The number of samplings per cell and cell-division step affects the phase-space sampling
procedure during the foam build-up. The value has to be large enough to fill the density
histograms used for the evaluation of the variance with sufficient statistical accuracy.
On the other hand, increasing this parameter to a value much larger than the average
number of training events contained in a cell will not improve the performance any further, 
as the sampling accuracy is limited in this case by the available number of training events in
the cells. The foam build-up time scales approximately linearly with the number of samplings.
The default value of 2000 is sufficiently large for optimal performance with all examples 
studied. For many cases with a small number of observables or small training-sample sizes, 
a reduced value of 500-1000 can be chosen without loss in performance.

\subsubsection{Number of bins}
Histograms are used to evaluate the variance across the cells projected on the cell axes.
Cell splits are only performed at the bin boundaries. The accuracy of the determination of
the division point increases with every iteration,
as the histograms are refined with respect to the base cell after each
division step.
For all examples studied, the dependence both of the performance and
the foam build-up time
on the number of bins was found to be very small. The default value
of 5 was found to be sufficient to achieve optimal performance.

\subsubsection{Gaussian kernel smoothing}
Foams with small number of cells and which are based on small training-sample sizes can
suffer from large cell-to-cell fluctuations leading to large 
discontinuities at the cell boundaries. A Gaussian smearing can be applied during the 
classification phase to reduce the effect of these discontinouities~\cite{kernel-cranmer}. 
In this case,
all cells contribute to the discriminant calculation for a given event, convoluted with
their Gauss-weighted distance to the event. The width parameter of the Gauss
function used for the smearing is set to the length of the sampling
box in each dimension (\texttt{VolFrac}).

\begin{figure}[htp!]
\centering
   \subfigure[Foam projection without kernel]{%
      \label{fig:foam-nokernel}
      \includegraphics[width=0.48\textwidth,clip]{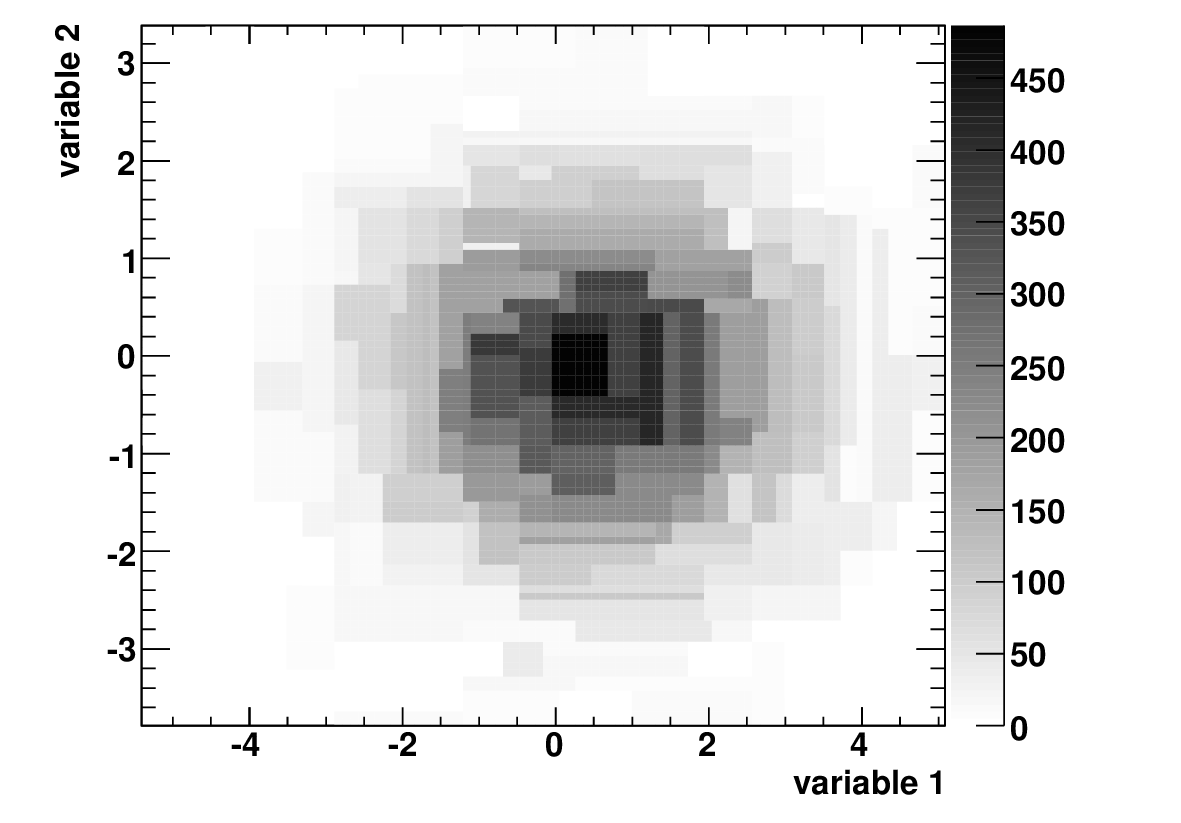}}
   \subfigure[Foam projection with Gaussian kernel]{%
      \label{fig:foam-kernel}
      \includegraphics[width=0.48\textwidth,clip]{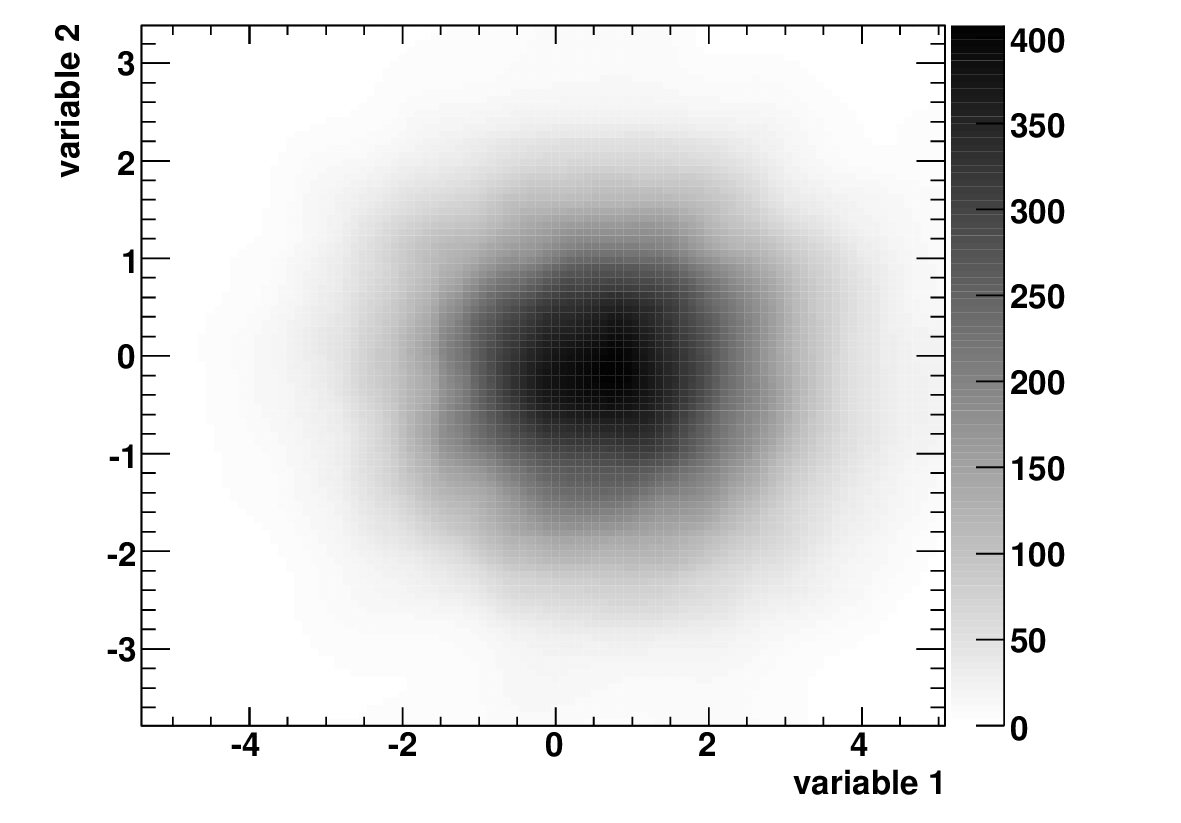}}
\caption[.]{Foam representations of a two-dimensional foam with 250 active cells
    for a Gaussian distribution with width 1.0 in each dimension and
    centred at (0.5, 0). The foam
    was created with 5000 signal and 5000 background training events. (a) shows the
    event-density distribution without using kernel weighting and (b)
    shows the distribution after smearing with a Gaussian kernel of
    width 0.33. The level of grey
    indicates the event density inside the corresponding cell. The
    foam boundaries are rescaled to the coordinate system of 
    the original observables.}
\label{fig:foam_kernel}
\end{figure}

Figure~\ref{fig:foam_kernel} shows the geometry
of a foam with 250 active cells and the reconstructed event density, based on 5000
signal and 5000 background training events generated according to a
two-dimensional Gaussian distribution with width 1.0 in each dimension 
and centred at (0.5,0). The reconstructed 
event densities with and without Gaussian kernel smearing are compared
in Fig.~\ref{fig:foam-nokernel}
and~\ref{fig:foam-kernel}. The width of the Gaussian kernel used for
the smearing corresponds to 0.33 in the units of the original distributions.
The improvement with kernel smearing is clearly visible. 
In most cases this procedure leads to an improved separation between signal and background. 

\begin{figure}[htp!]
\centering
\includegraphics[width=0.48\textwidth]{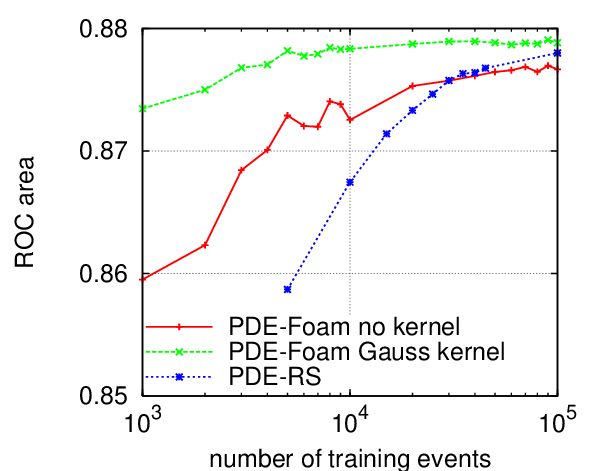}
\caption[.]{\texttt{PDE-Foam} performance as function of the number of training signal and background events for an example with
two-dimensional Gauss distributions with shifted mean values for signal and background. 
Shown is the performance for a foam with 250 active cells, both without and with Gaussian kernel 
smearing during the classification phase. The performance of the
original \texttt{PDE-RS} method is also shown for comparison.}
\label{fig:foamperf_kernel}
\end{figure}

Figure~\ref{fig:foamperf_kernel}
shows the performance as function of the number of signal and background training 
events\footnote{Here and in the following figures, the number of training events corresponds to the
individual sizes of both the signal and background samples. The actual total sample size is
therefore twice the number of events shown on the x-axis of the figures.}
for an example with
two-dimensional Gauss distributions and in comparison with the
original \texttt{PDE-RS} method.
Signal and background distributions have shifted means but identical widths
in this example\footnote{The 
distributions correspond to the example ``Bivariate uncorrelated Gaussian probability
densities'' defined and discussed in Ref.~\cite{pders-carli-koblitz}.}.
The foams contain 250 active cells and a cut on the minimum number of
events per cell was not applied. 
The Gaussian
kernel smearing improves the performance, in particular for
small training samples. For this example, it also exceeds the one of the
original \texttt{PDE-RS} method. However, the Gaussian kernel smearing
also largely increases the classification time.
The classification times obtained using training signal and
background samples of 100000 events each and testing signal and
background samples of 500000 events each 
were approximately 1.5~min for \texttt{PDE-RS},
1~min for \texttt{PDE-Foam} without Gaussian kernel smearing and 1~h
for \texttt{PDE-Foam} with Gaussian kernel smearing.

\section{Results}
The two main limitations of the original \texttt{PDE-RS} method are:
\begin{itemize}
\item The performance of the \texttt{PDE-RS} method increases only slowly with the size of the training
samples. For good results, typically large samples of the order of 500000 events are needed. 
\item Both the CPU time needed for classification and the memory consumption during classification
increase with the number of training events. \texttt{PDE-RS} is therefore considered to be a slowly responding
classifier for most applications.
\end{itemize}
In the following we present a comparison of the performance and CPU-time consumption
between \texttt{PDE-Foam} and the original \texttt{PDE-RS} method.
The results are shown for the example with five moderately 
correlated observables. Other examples have been studied and similar results were obtained.

\subsection{Performance}
\begin{figure}[htp!]
  \centering
  \subfigure[ROC area]{%
    \label{fig:performance-ROC}
    \includegraphics[width=0.48\textwidth,clip]{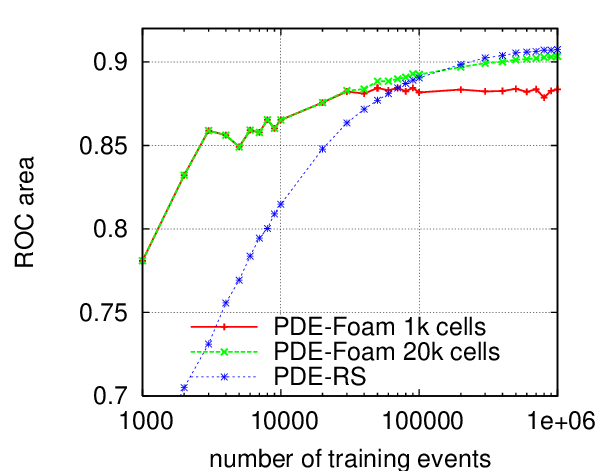}}
  \subfigure[Signal efficiency]{%
    \label{fig:performance-sigeff}
    \includegraphics[width=0.48\textwidth,clip]{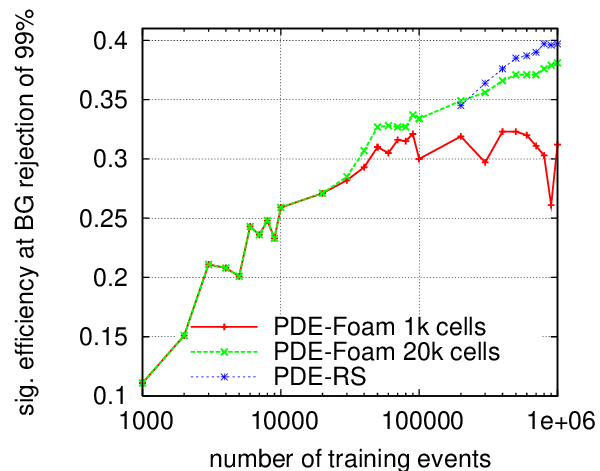}}
  \caption[.]{ROC area (a) and signal efficiency at a
    background rejection of 99\% (b) as function of the number of
    training signal and background events
    for the example with five moderately correlated observables constructed
    from Gaussian distributions. The performance for foams 
    with 1000 and 20000 active cells are compared to the performance of the original \texttt{PDE-RS} 
    method. The curve for \texttt{PDE-RS} in (b) starts from 200000 training events,
    as the method does not
    reach a background rejection of 99\% for smaller training-sample sizes.}
  \label{fig:performance}
\end{figure}

Figure~\ref{fig:performance} shows the estimator performance 
as function of the number of signal and background training events
for foams of 1000 and 20000 active cells, respectively. The left figure
displays the area under the ROC curve as a performance measure, while the
right figure shows the signal efficiency for a background rejection of 99\%.
The performance
of the original \texttt{PDE-RS} method is also shown. Single foams were built for these examples
with $\texttt{nSampl}=2000$ samplings, a sampling-box size of $\texttt{VolFrac}=1/30$
and a cut on the minimum number of events per cell of
$\texttt{Nmin}=100$. In case of \texttt{PDE-RS}, 
the sampling-box 
size was 1.2 in units of the original observables, corresponding to approximately 0.12
in normalised coordinates. 

For small training samples up to approximately 100000 events,
the foams perform significantly better than the original \texttt{PDE-RS} method. Apparently the
geometry of the foams is well adapted to the event distributions and the implicit averaging of the
event densities over the cell volumes leads to better performance than the
sampling with fixed box size performed by the original \texttt{PDE-RS}
method\footnote{A modified version of the original \texttt{PDE-RS} method is
  available within TMVA that allows to calculate the discriminant based on
  adaptive probe volumes and with kernel smearing. This can lead to
  improved 
  classification performance at the
  cost of an increased classification time. 
  Here we only consider the original
  \texttt{PDE-RS} implementation~\cite{pders-carli-koblitz}.}.
For training samples of less than
200000 events, the original \texttt{PDE-RS} method does not even reach a background rejection
of 99\%.

For very small training
samples of 30000 events and less, the foams with 1000 and 20000 cells behave identically, since
the cut on the minimum number of events per cell of 100 limits the effective number of final
cells to a value below 1000. 

For large training samples above 50000 events, the foam with 20000
cells performs better than the one with 1000, taking advantage of its finer granularity and
the increased statistical precision of the larger training samples. However, for
training-sample sizes of more than 200000 events, it does not quite reach the performance 
of the original \texttt{PDE-RS} method. For such large sample sizes, the local density estimates obtained with
the \texttt{PDE-RS} method by counting events in the vicinity of the events to be classified 
are more precise than the density estimates from counting events in foam cells
of finite granularity.

\subsection{CPU time}
\begin{figure}[htp!]
\centering
   \subfigure[Training time]{%
      \label{fig:training_time}
      \includegraphics[width=0.48\textwidth]{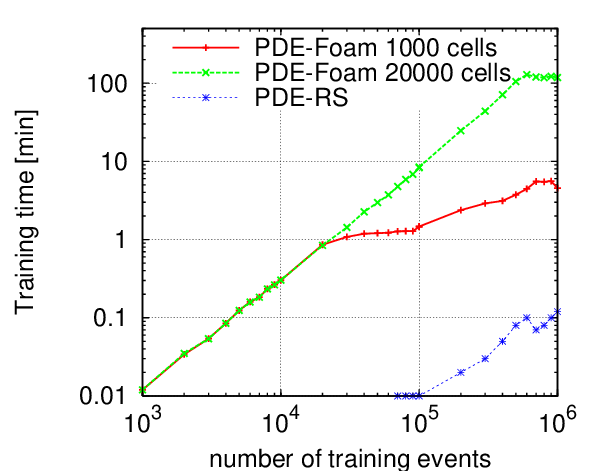}}
   \subfigure[Classification time]{%
      \label{fig:testing_time}
      \includegraphics[width=0.48\textwidth]{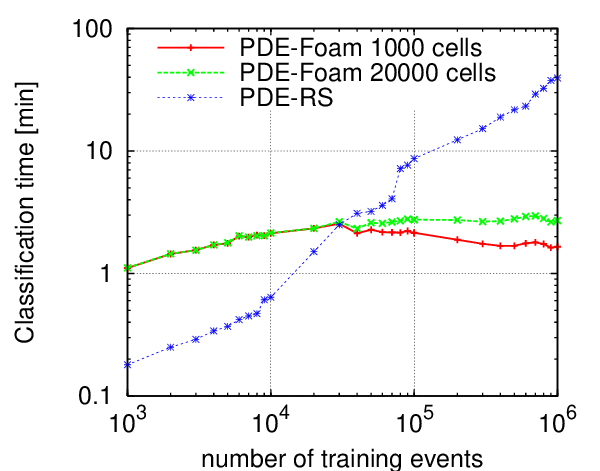}}
\caption[.]{CPU time for training (a) and classification of 500000 testing events (b) 
as function of number of training events 
for a model with five observables. The CPU time for foams 
with 1000 and 20000 cells are compared to the performance of the original \texttt{PDE-RS} 
method.}
\label{fig:time}
\end{figure}

Figure~\ref{fig:training_time} shows the training time as function of the number of training signal and background events 
for foams of 1000 and 20000 active cells, respectively, for the example and parameters described 
above\footnote{The values shown correspond to the CPU time spent on computers of the
CERN lxbatch computing cluster, running typically at 2.33~GHz.}.
The training time for the original \texttt{PDE-RS} method
is also shown. 
For \texttt{PDE-RS}, the training time consists only of the creation of the binary-search trees
used to store the training samples. For \texttt{PDE-Foam}, the training time is dominated by the repeated 
density sampling during the iterative build-up of the foam structure. Therefore the training time 
is larger than for the \texttt{PDE-RS} method. The training time 
for small training samples is identical for the foams with 1000 and 20000 cells, due to the
cut on \texttt{Nmin}.

Figure~\ref{fig:testing_time} shows the CPU time used for classification of 500000 signal and 500000 background
testing events
as function of the number of signal and background training events. For the foams, the classification time depends
mostly on the number of cells in the final foam and is almost 
independent of the number of training events. The slight variation with the number of training events
is due to the corresponding increase of the number of cells and due to slight
variations of the foam geometry. 
For the original \texttt{PDE-RS} method, on the
other hand, the classification time rises with the number of training events, due to 
the larger size of the binary trees. 
For $10^6$ signal and background training events each,
the classification time reaches approximately 40 minutes for \texttt{PDE-RS}, while for the foams with
20000 cells it is below 3 minutes.
On the other hand, for small training samples of
less than approximately 30000 events, the recursive reconstruction of
the foam geometry during classification takes longer than the density sampling within the
\texttt{PDE-RS} binary tree. 

\section{Reconstruction of event quantities}

The foam can be extended to reconstruct event quantities
(regression analysis). In this case $n_t$ target values depend on $d$ observables.
Two different methods have been implemented: 
the first method stores a single target value in every foam cell.
The second method saves the target values in further foam dimensions. Since
the first method can only be used if only one target is given, it is called
'Mono target regression'.  In order to do regression with multiple
targets one has to use the second method, called 'Multi target
regression'.  

In case of the mono-target regression, the density $\rho$ used during
the foam build-up phase, is given by the mean target
value $\langle t\rangle$ within the sampling box, divided by
the box volume (given by the \texttt{VolFrac} option):
\begin{equation}
\rho = \frac{\langle t\rangle}{\texttt{VolFrac}^d}
 \equiv
\frac{\sum_{i=1}^{N_b} t^{(i)}}{N_b \times \texttt{VolFrac}^d},
\end{equation}
where the sum goes over all events $N_{b}$ within the
sampling box and $t^{(i)}$ is the target value of the event
${\textbf{x}}^{(i)}$ ($i=1, \ldots, N_b$). During the foam build-up phase,
the relative variance of the density $\sigma_{\rho}/\langle\rho\rangle$
is minimised in the same way as described in 
section~\ref{foam-build-up-algorithm}.

After build-up of the foam, each cell is filled with the average target value.
During classification the target value is estimated for any
given event $\textbf{x}$ and is given as the content of the corresponding foam cell.

In case of multi-target regression, the $n_t$ target values are treated as additional
dimensions during the foam build-up. The density used for the foam build-up 
is estimated from the number of events in a box of fixed size in the $d+n_t$-dimensional 
phase space. The number of events contained in the volume of each cell of the final 
foam is stored with the foam. The target values for any given
event $\textbf{x}$ are estimated as the projections of the centre of the corresponding cell
onto the corresponding axes formed by the $n_t$ target values.

\begin{figure}[htp!]
\centering
   \subfigure[Foam geometry and target density]{%
      \label{fig:foam_reg}
      \includegraphics[width=0.48\textwidth]{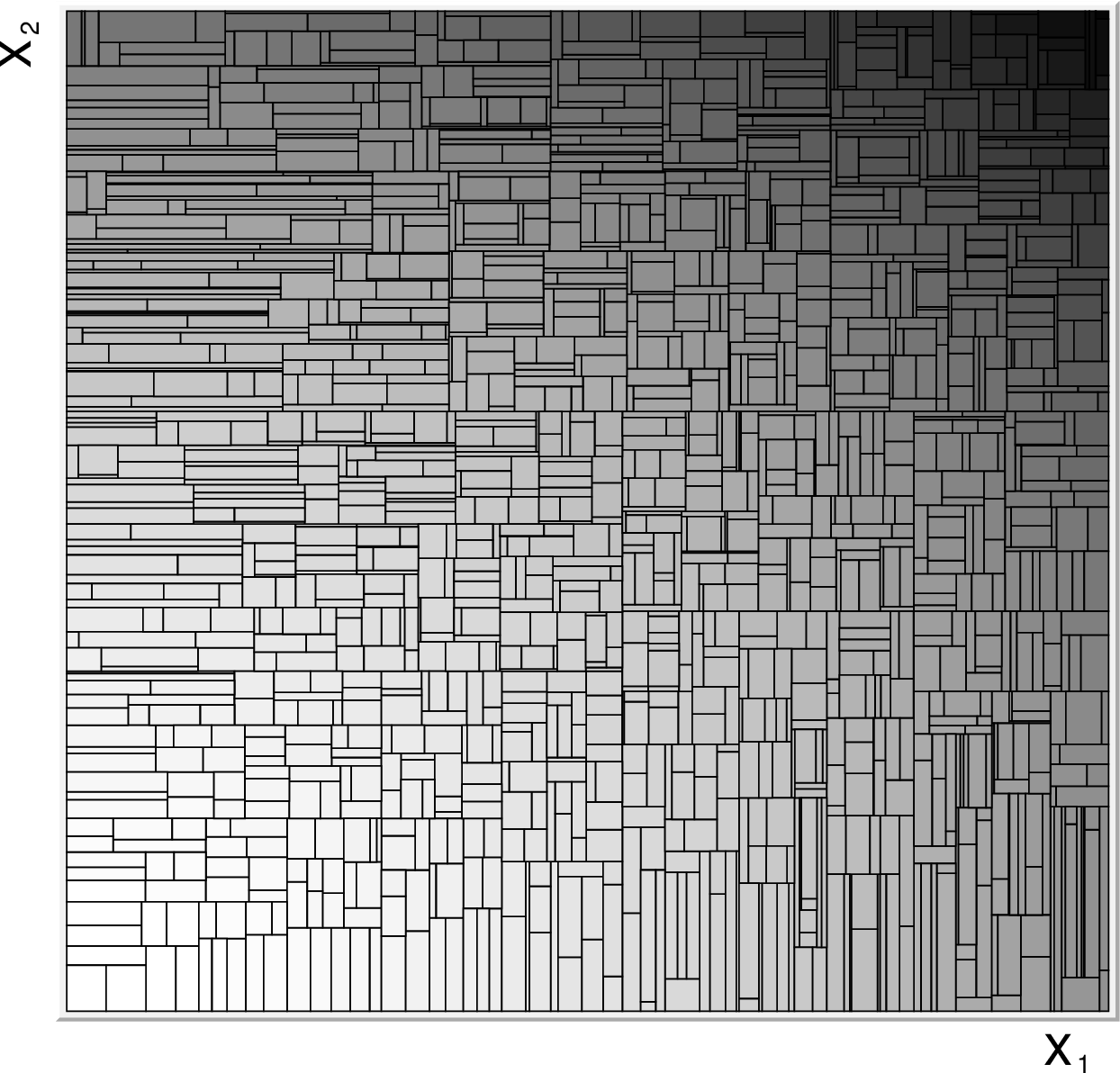}}
   \subfigure[Accuracy of event reconstruction]{%
      \label{fig:reg_accuracy}
      \includegraphics[width=0.48\textwidth]{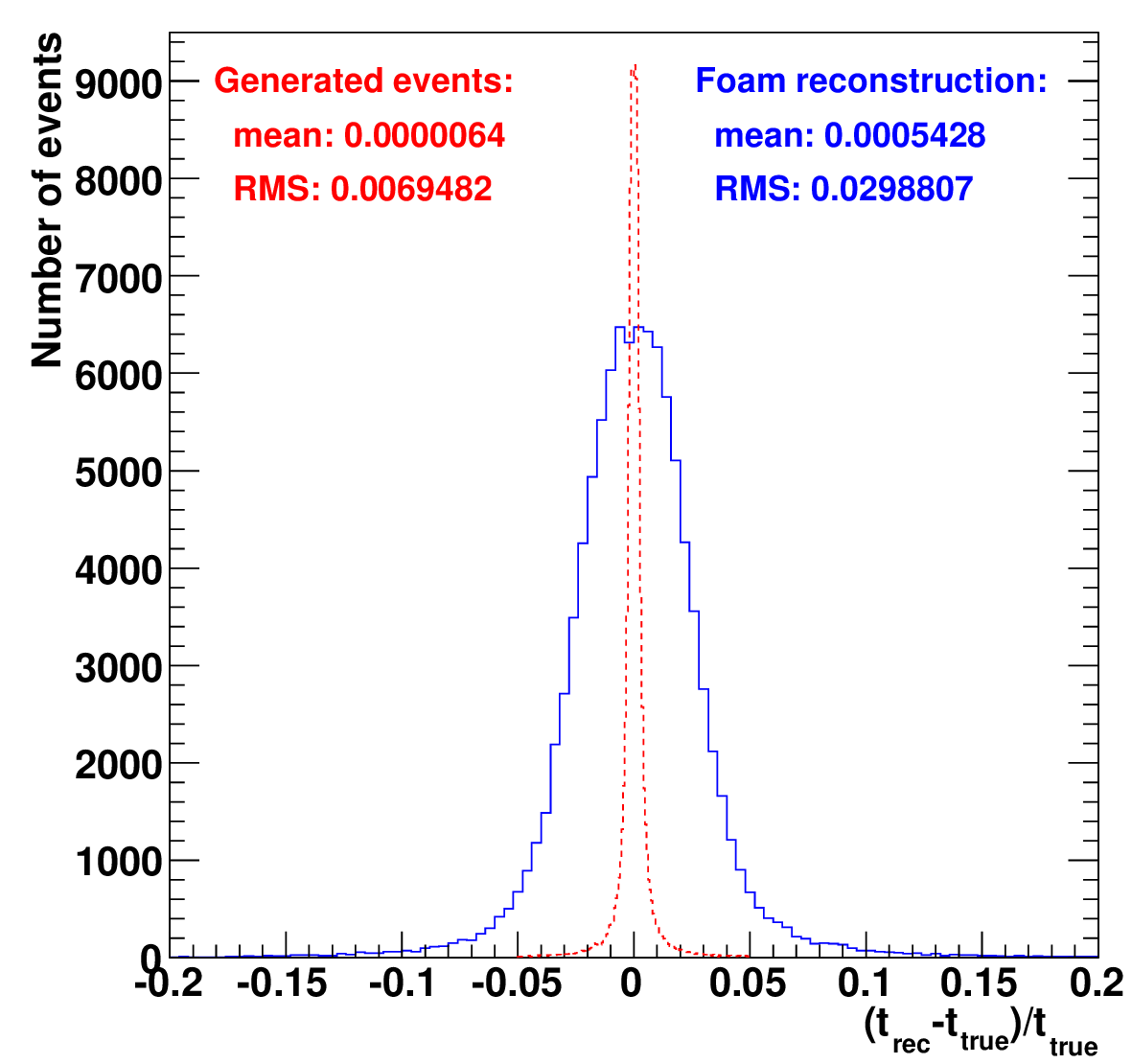}}
\caption[.]{Mono-target regression for a bivariate quadratic target function.
  (a) shows the geometry and target density of the resulting foam with 1500 active cells. 
  The level of grey
  indicates the target density inside the corresponding cell.
  In (b) the dashed line shows the relative difference between the
  generated events before and after adding the noise term and the
  solid line shows the accuracy of the event reconstruction with \texttt{PDE-Foam}. }
\label{fig:regression}
\end{figure}

Figure~\ref{fig:foam_reg} shows the geometry and the target density for a mono-target foam 
with 1500 active cells, calculated for an example with two observables and 
a quadratic dependence of the target value $t$ on two uniformly distributed observables,
$\textrm{x}_1$ and $\textrm{x}_2$:
\begin{equation}
t=a\cdot \textrm{x}_1^2+b\cdot \textrm{x}_2^2+c+\Delta t,
\end{equation}
where $a$, $b$ and $c$ are constant and $\Delta t$ is a small random
number simulating noise. The accuracy of the event reconstruction with
this foam is shown in Fig.~\ref{fig:reg_accuracy}, where the relative 
difference between the reconstructed and true target value is
displayed.
The mean value is reconstructed with an accuracy of approximately 0.5 per mille. The 
RMS of the distribution is about 3\%.
Also shown in the figure is the relative difference between the generated events
before and after adding the noise term. The width of this distribution
is approximately 0.7\%. It can be considered as the optimal value for
the accuracy of the event reconstruction. 

\section{Conclusions}
A new method for multi-variate analysis, \texttt{PDE-Foam}, has been developed. It combines the adaptive
binning algorithm of the \texttt{Foam} method so far only used for Monte-Carlo event generation
with probability density estimation based on range searching.
\texttt{PDE-Foam} has been implemented within the TMVA package for multi-variate analysis.

We demonstrated that the default set of foam build-up parameters leads to robust
results for the various examples studied and we gave guidance for further parameter optimisation.
We showed that the performance of \texttt{PDE-Foam} exceeds the classification
performance of the original \texttt{PDE-RS} method for small training
samples. Furthermore, it leads to largely reduced classification time.
Both the classification time and the
memory consumption are independent of the number of training
events. The main limitations of 
the original \texttt{PDE-RS} implementation have therefore been overcome.

In addition to event classification we have implemented a method to reconstruct event 
quantities with \texttt{PDE-Foam}. 

\section*{Acknowledgements}
The authors would like to thank Stanislaw Jadach for many explanations and enlighting discussions 
concerning the \texttt{Foam} algorithm. We are grateful to Andreas H\"ocker for fruitful discussions in the development
phase and his support in implementing the \texttt{PDE-Foam} within the TMVA framework.
We would also like to thank the other main TMVA developers and in particular J\"org Stelzer
and Helge Voss for their help. We are very grateful to Birger Koblitz and Yair Mahalalel
for their help with the \texttt{PDE-RS} method. We also benefited a lot from a discussion with
Fred James.

\end{document}